# Stainless steel in an electronically excited state

Nikita Medvedev[1,2,*]

1) Institute of Physics, Czech Academy of Sciences, Na Slovance 1999/2, 182 00 Prague 8, Czech Republic
2) Institute of Plasma Physics, Czech Academy of Sciences, Za Slovankou 3, 182 00 Prague 8, Czech Republic

## Abstract

Understanding the non-equilibrium behavior of stainless steel under extreme electronic excitation remains a critical challenge for laser processing and radiation science. We employ a hybrid framework integrating density-functional tight binding, transport Monte Carlo, and Boltzmann equations to model austenitic stainless steel ($Fe_{0.5875}Cr_{0.25}Mn_{0.09}Ni_{0.07}C_{0.0025}$) under ultrafast irradiation. The developed approach uniquely bridges atomic-scale electronic dynamics and mesoscale material responses, enabling the quantitative mapping of electron-temperature-dependent properties (electronic heat capacity, thermal conductivity, and electron-phonon coupling) up to the electronic temperatures $T_e$~25,000 K. Two distinct lattice disordering mechanisms are identified: nonthermal melting at $T_e$~11,500 K (the dose ~1.9 eV/atom), where the lattice collapses on sub-picosecond timescales without atomic heating driven by electronic excitation modifying the interatomic potential; and thermal melting (at ~0.45 eV/atom), induced by electron-phonon coupling on picosecond timescales. The derived parameters enable predictive modeling of stainless steel under extreme conditions, with implications for laser machining and radiation-resistant material design.

## I. Introduction

Austenitic stainless steel is a multi-principal element alloy based on iron with added chromium, nickel, carbon, and other elements such as molybdenum or titanium, depending on the particular steel grade [1]. It is widely used due to its resistance to corrosion, good formability, non-magnetic properties, and mechanical stability in a wide temperature range. Invaluable applications include medical uses such as nondegradable implants[2] and surgical instruments [3]; in engineering and construction it is used, for example, in corrosive environments and aerospace [4]; in chemistry and petrol industry it is utilized for piping, vessels, and in chemically corrosive environments [5]. Stainless steel is also applied in radiation harsh environments, such as nuclear reactors [6,7], particle accelerators [8], and free-electron laser beam monitors [9]. It is, thus, essential to understand the fundamental processes taking place in the response of the material to irradiation.

Damage in materials induced by photon irradiation differs, depending on the parameters of the laser pulse: its photon energy, pulse duration, intensity, etc. [10,11]. Ultrafast laser pulses, with sub-picosecond duration, induce specific transient effects and corresponding observable material modification, not present in irradiation by nanosecond or longer pulses [10,12]. There is a typical sequence of processes leading to material damage under ultrafast laser irradiation: first, photoabsorption during the laser pulse promotes electrons into high energy states, driving the electronic ensemble out of equilibrium. Depending on the photon energy, conduction/valence

* Corresponding author: email: nikita.medvedev@fzu.cz, ORCID: 0000-0003-0491-1090

band electrons may absorb photons (typical for near-infrared and optical lasers), or atomic core-shell electrons may be excited (for X-ray photons) [13,14]. Such electrons then scatter with other electrons, performing secondary cascading. For high electron energies, they may excite secondary electrons from core shells – impact ionization process – or scatter on valence and conduction band electrons [14,15]. At the same time, they scatter quasi-elastically on the atomic lattice *via* the electron-ion, or electron-phonon, coupling [13,16,17].

Simultaneously, core-shell holes may decay *via* the Auger channel, exciting new secondary electrons, or fluorescence emitting secondary photons [18]. All the secondary electrons then join the kinetics of the photo-electrons, participating in the cascading, which eventually leads to electron thermalization: establishment of the equilibrium Fermi-Dirac distribution of electrons [19,20]. At this stage, typically occurring at femtosecond time-scales, the electrons acquire thermodynamic temperature different from that of the atomic ensemble [13,20].

The influence of the high-temperature electrons on the atomic lattice is two-fold: they exchange the kinetic energy with the atomic system (*via* the abovementioned electron-phonon coupling), and they also exchange the potential energy, by modifying the interatomic potential, which may lead to a phase transition even without significant heating of the atoms [21]. The most famous example of the latter is the nonthermal melting [22,23]; in chemistry, it is known as photo-induced bond breaking [24]. Nonthermal melting commonly occurs in covalent materials but is rare in metals [21,25,26].

The thermal effects – heating *via* the electron-phonon coupling and eventual melting upon overcoming the melting point – are often described theoretically in the framework of the two-temperature model [13,16,27]. In this model, the electronic system and the atomic (phononic) systems are described with a system of coupled thermodynamical equations, whose parameters depend on the electronic and atomic temperatures. The current state of the art in the large-scale modeling of the laser irradiation of metals is the two-temperature molecular dynamics, which couples the thermodynamic (continuum) description of electrons with the molecular dynamic (atomistic) description of the atomic system [26,28,29]. In such simulations, the key parameters are the electronic-temperature-dependent electronic heat capacity, electronic heat conductivity, and electron-ion (electron-phonon) coupling [16,17].

This forms the goal of the current work: to evaluate the electron-temperature-dependent thermodynamic parameters of the electronic system, and to establish whether the nonthermal melting is possible in stainless steel.

## II. Model

To study material under ultrafast irradiation, the XTANT-3 simulation toolkit is applied [30]. It is a multiscale hybrid model, combining a few simulation approaches to trace all the essential effects of irradiation: tight-binding molecular dynamics for the atomic system, and Boltzmann equation and transport Monte Carlo simulation for the electrons.

The photoabsorption, the nonequilibrium electron cascades, and Auger-decays of core holes (if any) are modeled with the event-by-event transport Monte Carlo model. The photoabsorption cross sections, the ionization potentials of core shells, and Auger-decay times are taken from the EPICS2023 database [31] (see photon attenuation lengths in the Appendix). The impact ionization cross section of fast electrons (with energies above the cut-off of 10 eV) is calculated with the complex-dielectric function formalism, applying the single-pole method [32] (see electron inelastic mean free paths in the Appendix). Quasi-elastic scattering of fast electrons is described with the modified Molier cross-section [33].

Slow electrons (with energies below the cut-off) are traced with the help of the Boltzmann equation [20]. The electron-electron thermalization is modeled within the relaxation time approximation, whereas for the electron-ion (or electron-phonon) scattering the dynamical coupling approach is used, see below [17]. These electrons are populating the transient energy levels (band structure), evolving in time. The evolution of the electronic energy levels is described with the transferrable tight binding method [15,34,35]. DFTB method with the PTBP tight binding parameterization (with the $sp^3d^5$ linear combination of atomic orbitals (LCAO) basis set) is used to evaluate the electronic Hamiltonian[36]. Its diagonalization produces the electronic energy levels (molecular orbitals, band structure), and the gradients serve to evaluate the interatomic forces [34].

The atomic trajectories are traced with the molecular dynamics (MD) method. The interatomic forces are calculated *via* the gradients of the potential energy, obtained from the tight-binding Hamiltonian, and the transient electronic populations (electronic distribution function modelled with the Boltzmann equation) [15,34,35]. Since the changes in the electronic distribution function directly affect the interatomic potential, the method is capable of reproducing nonthermal melting effects [15,37]. Martyna-Tuckerman $4^{th}$ order algorithm is used with the timescale of 1 fs or smaller [38].

Austenitic stainless steel with the chemical composition of $Fe_{0.5875}Cr_{0.25}Mn_{0.09}Ni_{0.07}C_{0.0025}$ may be exactly reproduced using a simulation box containing 400 atoms (235 atoms of Fe, 100 Cr, 36 Mn, 28 Ni, 1 C). For the tight binding parameterization used, defining the interatomic potential, the equilibrium dimensions of the box are 18.45 x 18.4 x 14.76 $Å^3$, which produces the material density of 7.3 $g/cm^3$ (slightly lower than the experimental values of ~7.5-7.9 $g/cm^3$; for multi-elemental transferable tight binding methods, such differences are typical and expected, because the parameters are not specifically fitted to reproduce one particular material [39]).

To initialize the simulation, atoms are randomly placed on the fcc-structure grid inside the simulation box, creating a homogeneously mixed alloy. For each simulation, a new random placement of atoms is performed, and results are averaged over a few realizations of the initial structure to ensure the independence of the results of the particular atomic placement. Ten simulations are used for the evaluation of the electronic heat capacity and conductivity, whereas 40 simulation runs are used for calculations of the electron-phonon coupling parameter.

For the dynamical damage simulations, the system is equilibrated at room temperature before the arrival of the laser pulse. All the theoretical and numerical details of XTANT-3 may be found in Ref. [30].

Having the model for the electronic energy levels (band structure), wave functions, electronic populations, and atomic structure and dynamics, the following methods are used to evaluate the electronic-temperature-dependent parameters. The electronic heat capacity is calculated as follows [30]:

$$C_e(T_e) = \frac{1}{V_0}\sum_i \frac{\partial f_e(E_i)}{\partial T_e}(E_i - \mu(T_e)), \qquad (1)$$

here $V_0$ is the volume of the simulation box; $E_i = \langle i|H\{\boldsymbol{R}_{jk}\}|i\rangle$ are the electronic energy levels obtained *via* diagonalization of the transient Hamiltonian $H\{\boldsymbol{R}_{jk}\}$ dependent on the relative position of all the atoms in the simulation box; $\mu(T_e)$ is the electronic chemical potential, dependent on the electron temperature $T_e$; and $f_e(E_i)$ is the electronic distribution function (fractional populations of electrons) on the energy levels (Fermi-Dirac distribution in thermal equilibrium).

The electronic heat conductivity is evaluated *via* two terms: the electron-atom (electron-phonon, $\kappa_{e-a}$) and electron-electron ($\kappa_{e-e}$) scattering contributions combined according to Matthiessen's rule [40,41]:

$$\kappa_{tot}(T_e) = \left(\frac{1}{\kappa_{e-a}(T_e)} + \frac{1}{\kappa_{e-e}(T_e)}\right)^{-1} \tag{2}$$

The electron-phonon part is calculated within the Kubo-Greenwood formalism [42]:

$$\kappa_{e-a}(T_e) = L_{22} - T_e \frac{L_{12}^2}{L_{11}} \tag{3}$$

Where the Onsager coefficients are [42]:

$$L_{ij} = -\frac{(-1)^{i+j}}{V_0 m_e} \sum \frac{df}{dE_k} (E_k - \mu(T_e))^{i+j+2} \, |\langle k|p|k'\rangle|^2 \tag{4}$$

with $m_e$ being the free electron mass, and the momentum matrix element calculated from the tight-binding Hamiltonian according to Ref. [43].

The electron-electron contribution is calculated analogously, with the matrix element replaced with the scattering cross-section from the Monte Carlo module of XTANT-3 [44].

The k-space grid used for evaluation of the DOS, electronic heat capacity and conductivity, are 7x7x7 points in the Monkhorst-Pack grid[45].

The electron-ion (electron-phonon) coupling parameter, $G(T_e, T_a)$, dependent on both the electronic and atomic temperatures ($T_e$ and $T_a$, correspondingly), is obtained with the nonperturbative dynamical coupling approach [17]:

$$G(T_e, T_a) = \frac{1}{V_0(T_e - T_a)} \sum_{i,j} E_j I_{e-a}^{ij} \tag{5}$$

The scattering integral, $I_{e-a}^{ij}$, is calculated as[17]:

$$I_{e-a}^{ij} = w_{ij} \begin{cases} f(E_i)\left(2 - f(E_j)\right) - f(E_j)(2 - f(E_i))e^{-E_{ij}/T_a}, \text{for i > j} \\ f(E_j)(2 - f(E_i))e^{-E_{ij}/T_a} - f(E_i)\left(2 - f(E_j)\right), \text{otherwise} \end{cases} \tag{6}$$

The scattering probability is obtained *via* the wave-function overlap on two consecutive MD steps $t = t_0 + \delta t$ (*via* the basis coefficients in the LCAO basis set within the tight binding Hamiltonian, $\psi_i = \sum_\alpha c_{i,\alpha}\varphi_\alpha$) as[17]:

$$w_{ij} \approx \frac{4e}{\hbar \delta t^2} \sum_{\alpha,\beta} \left|c_{i,\alpha}(t) c_{j,\beta}(t_0) S_{i,j}\right|^2 \tag{7}$$

Here $S_{\alpha,\beta}$ is the overlap matrix [34]. The dependence on the atomic temperature is implicit, as it enters dynamically in the MD simulation run [17,46].

All the illustrations of the atomic snapshots are prepared with the help of OVITO[47].

## III. Results

### a. *Electronic properties*

We start with the calculation of the density of states (DOS) of stainless steel. XTANT-3 calculates the orbital-resolved DOS for each element, based on the LCAO basis set [48]. Figure 1 shows the DOS compared to that calculated using density-functional theory in Ref.[49]. The differences between XTANT-3 and Ref.[49] calculations are, first, that Ref.[49] simulated $Fe_{73}Cr_{21}Ni_{14}$, a different composition from ours – e.g., it can be seen in particular peaks coming from Mn in our case. Second, the slightly lower density of the material here makes the peaks in the DOS narrower. Considering these differences, the agreement with the calculation from Ref.[49] is reasonable.

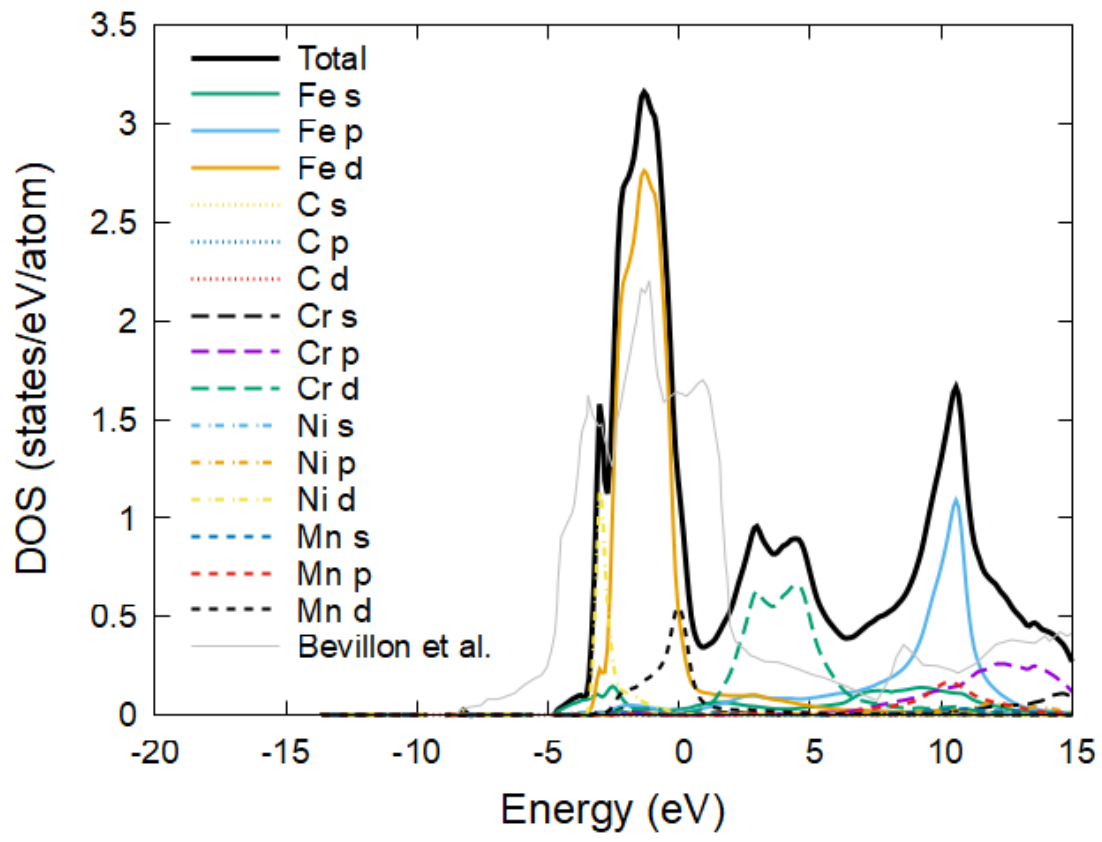

*Figure 1. The calculated total density of states (DOS) of unexcited stainless steel (fcc), and its element and orbital-resolved contribution, compared to the total DOS from Ref.* [49] (thin grey line). *The energy is counted from the Fermi level.*

The electronic chemical potential as a function of the electronic temperature in stainless steel is shown in Figure 2. The chemical potential is rising with the increase of the electronic temperature, qualitatively agreeing with Ref.[49]. The minor differences, again, are due to the different composition and density of the modeled material.

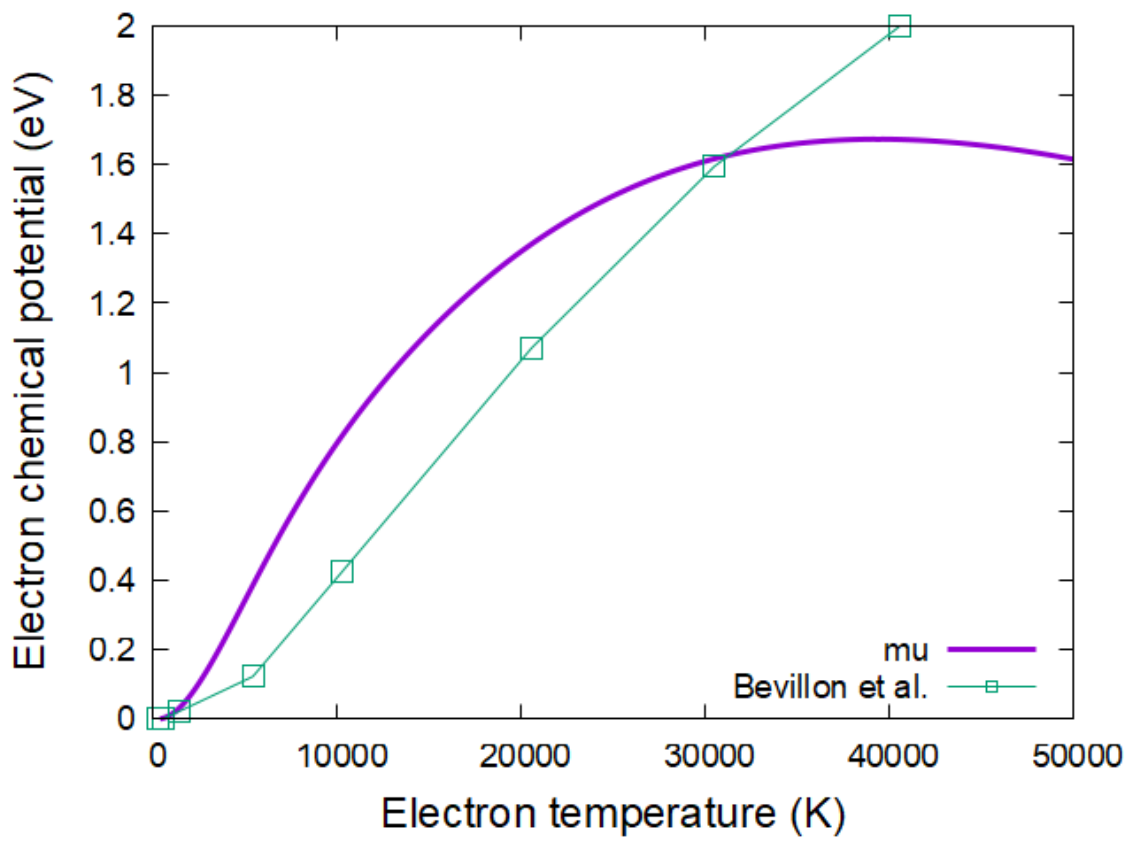

*Figure 2. The electronic chemical potential in stainless steel (fcc), compared to that from Ref.* [49]. *The energy is counted from the Fermi level.*

The electronic heat capacity as a function of the electronic temperature (calculated with the help of Eq.(1)) is shown in Figure 3. There, again, the overall agreement with that reported in Ref.[49] is observed. The difference has the maximum at the electronic temperature of $T_e$~30,000 K reaching ~30%.

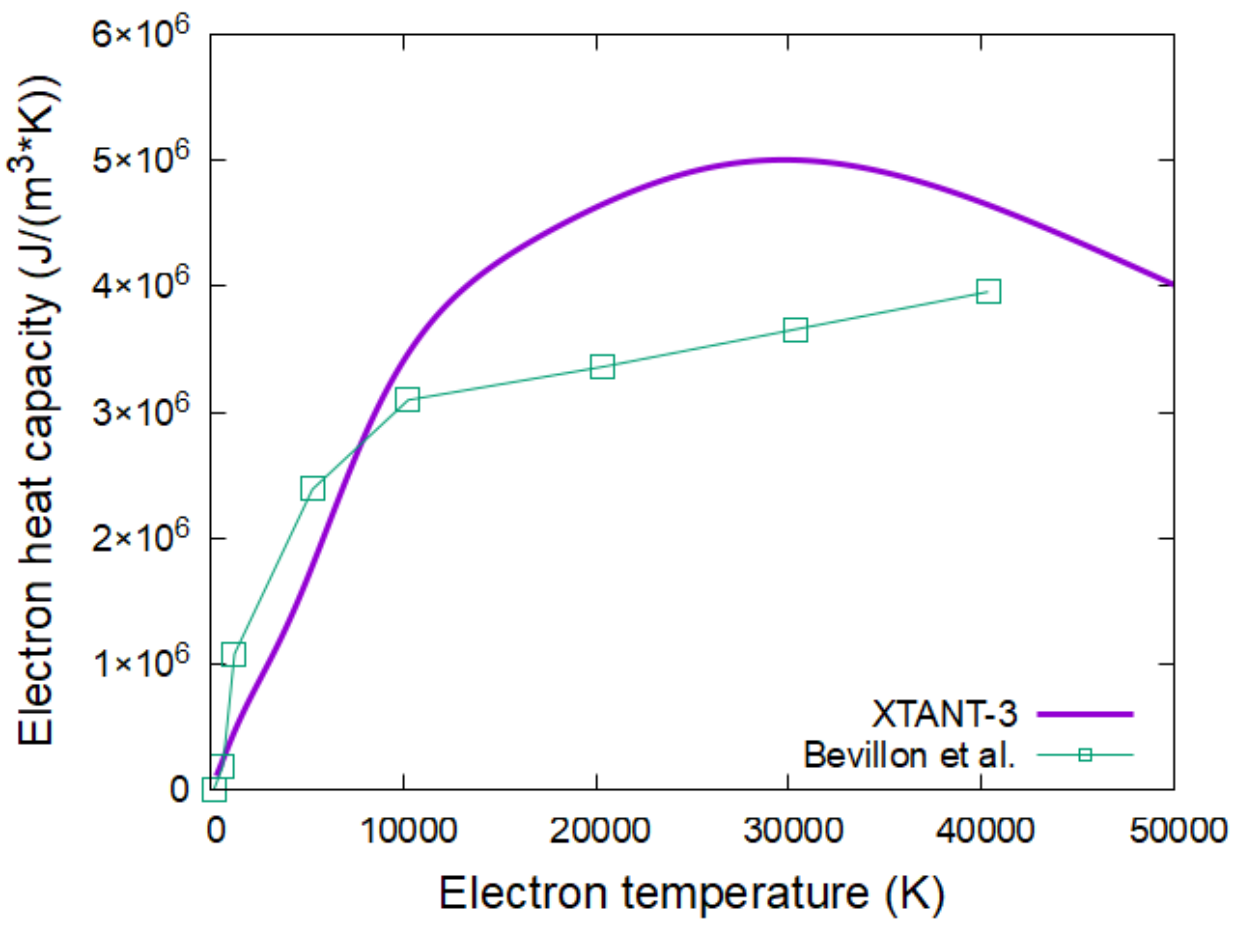

*Figure 3. The electronic heat capacity in stainless steel (fcc), compared to that from Ref.* [49].

The electronic heat conductivity (Eqs.(2-4)) is shown in Figure 4. The total electronic heat conductivity is similar to that from Ref.[49] at the electronic temperatures below $T_e$~20,000 K, above which the difference grows. Overall, the heat conductivity calculated with the DFT-based Kubo-Greenwood method tends to follow the shape of the electron-phonon contribution alone, since the electron-electron contribution is not included there (see a discussion in [44]). At high electronic temperatures, the electron-electron contribution starts to dominate, lowering the total electron heat conductivity [44,50]. This trend is visible in the XTANT-3 calculation in Figure 4.

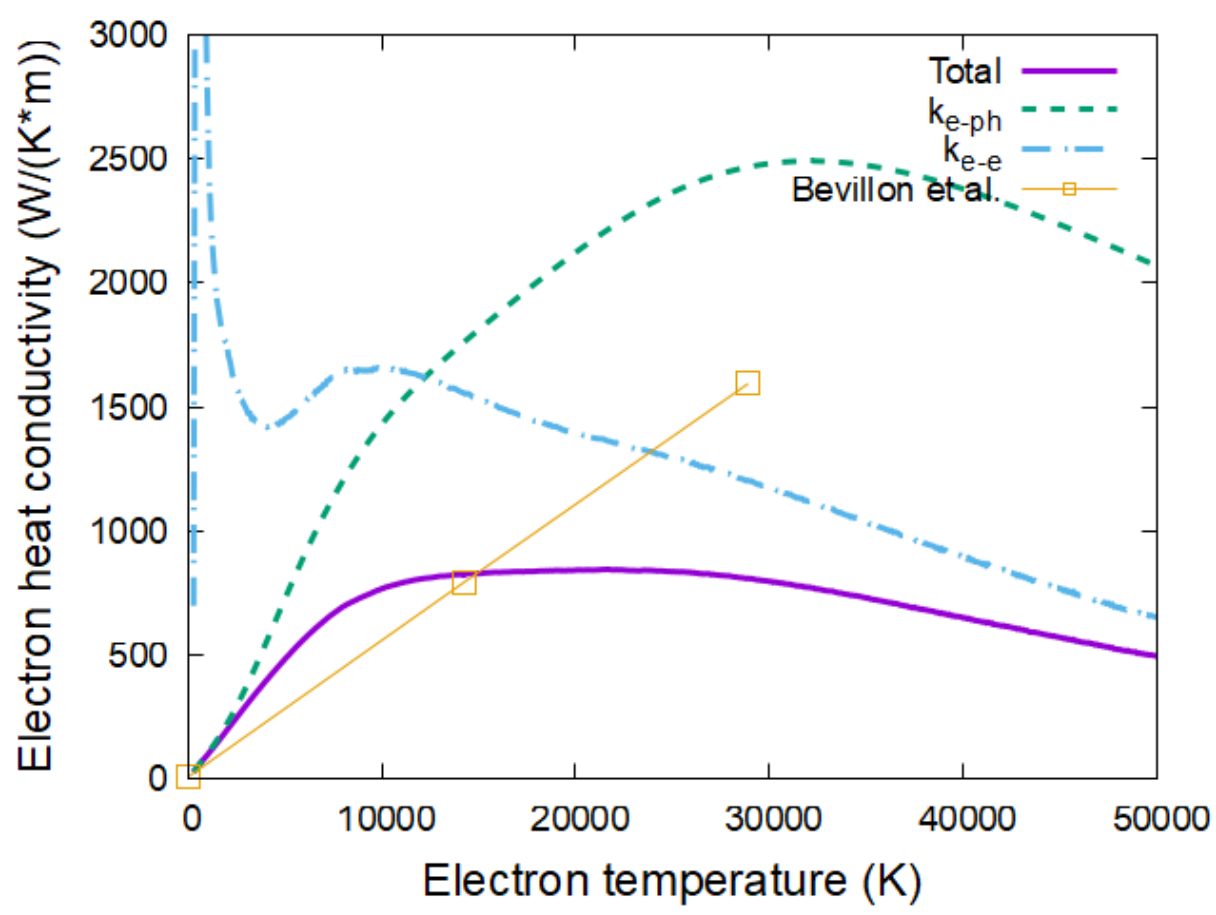

*Figure 4. The calculated electronic heat conductivity in stainless steel (fcc), and its electron-atom (electron-phonon) and electron-electron contributions, compared to that from Ref.* [49].

The electron-ion (electron-phonon) coupling is shown in Figure 5 for room atomic temperature. In XTANT-3, it is calculated with the dynamical coupling approach (Eqs.(5-7)), and thus could not be extended to temperatures above $T_e$~25,000 K [17]. The difference from the coupling from Ref.[49] is large – the main reason for it is that the Eliashberg formalism with Wang extension used in Ref.[49] typically overestimates the coupling parameter, see discussion in [17] and [51]. Additionally, a lower density of the material modeled with XTANT-3 can also lower the coupling parameter slightly[17].

It was noted in Ref.[52] that the electron-phonon coupling parameter may be sensitive to particular tight-binding parameterization used, which may also contribute to the difference from Ref.[49]. Ultimately, the calculated coupling parameter requires experimental validation, especially at high electronic temperatures.

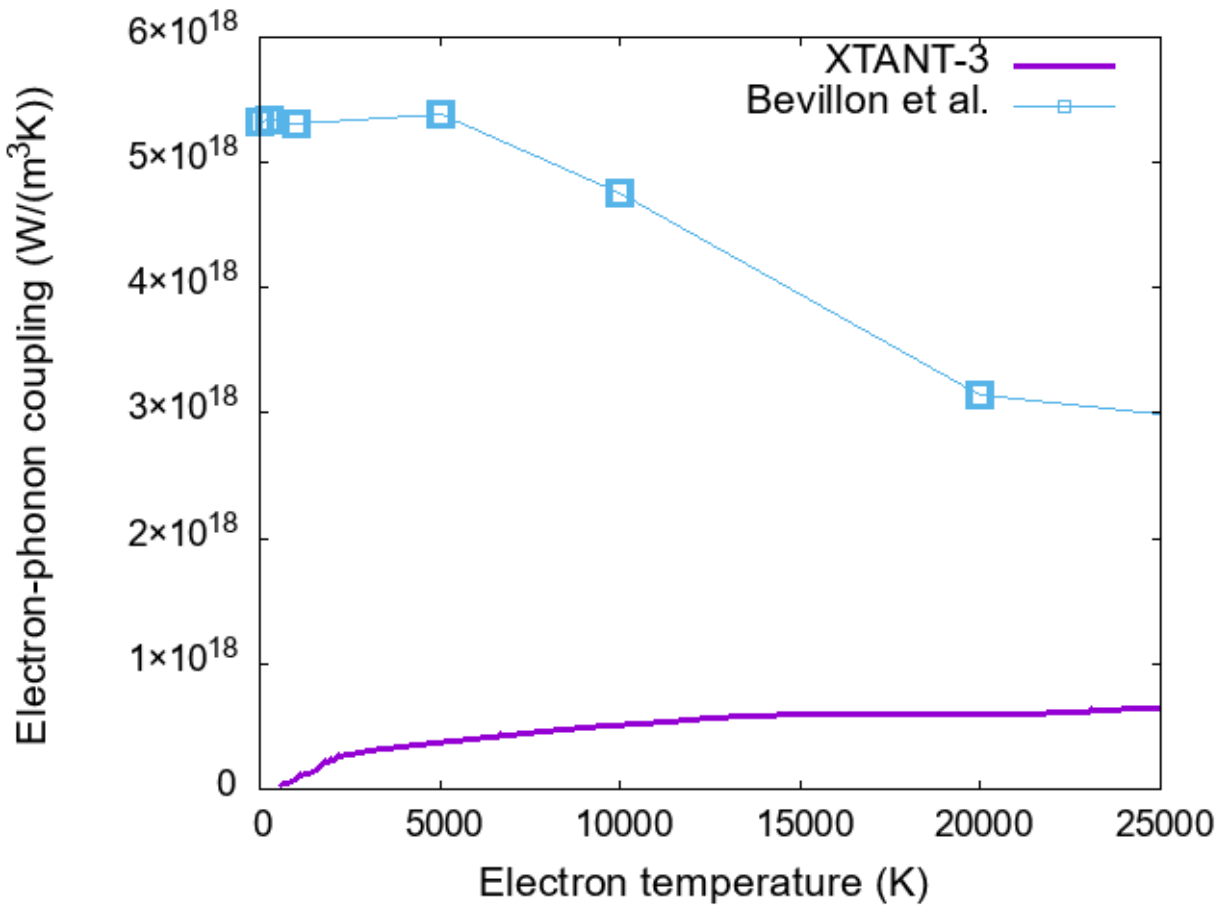

*Figure 5. The electron-phonon coupling in stainless steel (fcc) calculated with the nonperturbative dynamic approach in XTANT-3 at room atomic temperature ($T_a$ = 300 K), compared to that calculated with the Eliashberg formalism in Ref.* [49].

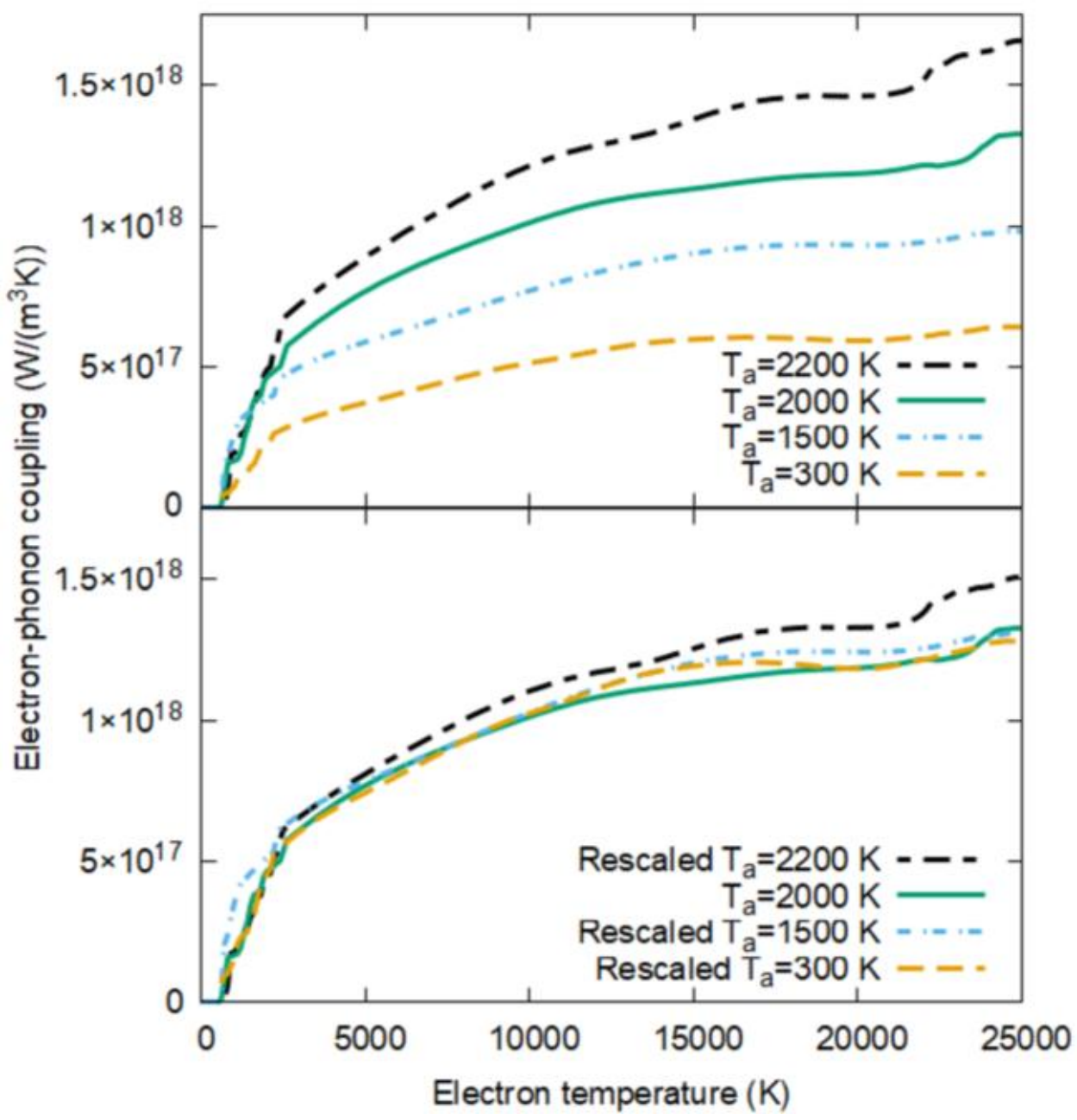

*Figure 6. (Top panel) The electron-phonon coupling in stainless steel (fcc) calculated with the nonperturbative dynamic approach in XTANT-3 various atomic temperatures. (Bottom panel) The same, but data rescaled with the linear coefficient (Eq.(8)) to match the data at $T_a$=2000 K.*

XTANT-3 calculation of the dynamical coupling at elevated atomic temperatures shows a linear increase of the overall curves, similar to the previously reported dependence in other materials [17,46]. Figure 6 demonstrates that, using the room-temperature curve as a base ($T_{room}$=300 K), the electron-phonon coupling parameter at higher atomic temperatures may be approximated by the linear rescaling:

$$G(T_e, T_a) \approx \alpha \frac{T_a}{T_{room}} G(T_e, T_a = T_{room}) \tag{8}$$

where the scaling coefficient is $\alpha = 0.3$. This linear dependence on the atomic temperature holds up to the onset of melting.

## *b. Ultrafast damage*

A sequence of dynamical simulations is performed to identify at what deposited dose the damage onsets. To model the damage formation, the laser pulse of 10 fs duration (full width at the half maximum, FWHM, of a Gaussian pulse), 30 eV photon energy, and various deposited doses are used. The damage onset is identified to take place at the dose of ~0.45 eV/atom, which coincides with the melting point of stainless steel, indicating that the damage is thermal. Above this dose, the material melts, with the iron lattice disordering at the picosecond timescale, see an example in Figure 7 for the dose of 1 eV/atom. The system disorders at the time of about 1.5-2 ps after the irradiation. Carbon atoms are very mobile, while other elements displace later [53,54]. A complete disorder takes place when the iron sublattice loses its stability.

During the 2 ps after the laser pulse arrival, the electronic and atomic temperatures did not equilibrate yet, see Figure 8. The melting point of stainless steel (~1600-1800 K) is reached by the time of ~1 ps. After that, it takes some time for the atomic system to disorder (cf. Figure 7).

Note that most elements have temperatures close to the average atomic temperature, with two exceptions. First, the carbon temperature oscillates strongly, because there is only a single carbon atom in the simulation box.

Second, the manganese subsystem temperature exhibits a spike just after the irradiation. It takes place before any significant electron-phonon coupling transfers the kinetic energy from excited electrons to atoms. Instead, manganese atoms accelerate nonthermally, experiencing the modification of the interatomic potential due to the electronic excitation (as discussed in detail in [55]). The effect of various subsystems responding differently to electronic excitation was previously reported in insulating materials [56], but not in metal alloys. Nonthermal acceleration suggests that nonthermal melting is possible in stainless steel.

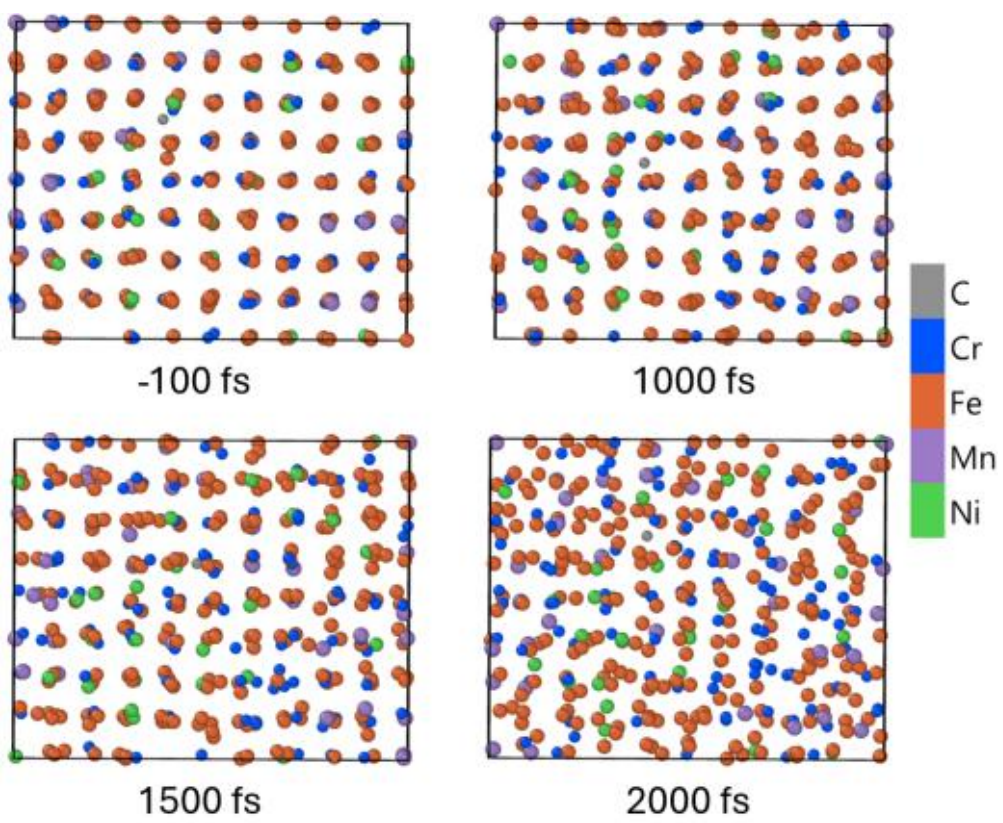

*Figure 7. Atomic snapshots of the supercell of stainless steel (fcc) after irradiation with 1 eV/atom absorbed dose, 10 fs FWHM, 30 eV photon energy, modeled with XTANT-3.*

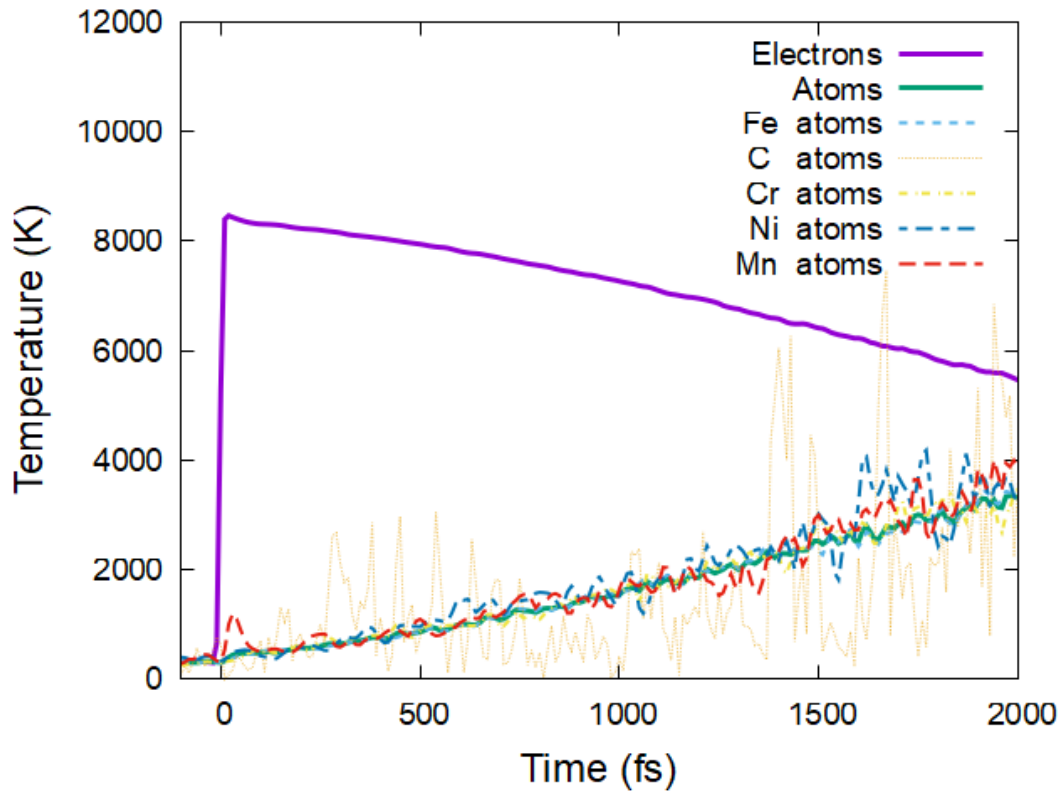

*Figure 8. Atomic and electronic temperatures in stainless steel (fcc) irradiated with 1 eV/atom absorbed dose, 10 fs FWHM, 30 eV photon energy, modeled with XTANT-3. Atomic temperatures of various species are also shown.*

To study this possibility, a separate set of simulations within the Born-Oppenheimer approximation was performed, which excludes the electron-phonon coupling. Eliminating the atomic heating *via* the non-adiabatic energy exchange allows us to identify a contribution of purely nonthermal melting.

In such a simulation setup, stainless steel starts melting at doses of about ~1.9 eV/atom, see an example in Figure 9. This is purely nonthermal melting, as the atomic system keeps at near-room temperature, except for a similar transient spike in the Mn subsystem, Figure 10. Despite the cold atomic system, it disorders at sub-picosecond timescale: the interatomic potential flattened due to electronic excitation, allowing atoms to overcome the potential barriers and melt [23,37,57,58]. After that, the atomic temperature rises slightly due to ongoing phase transition (nonthermal acceleration, a similar effect was observed, e.g., in diamond [15], and other materials [55,56]).

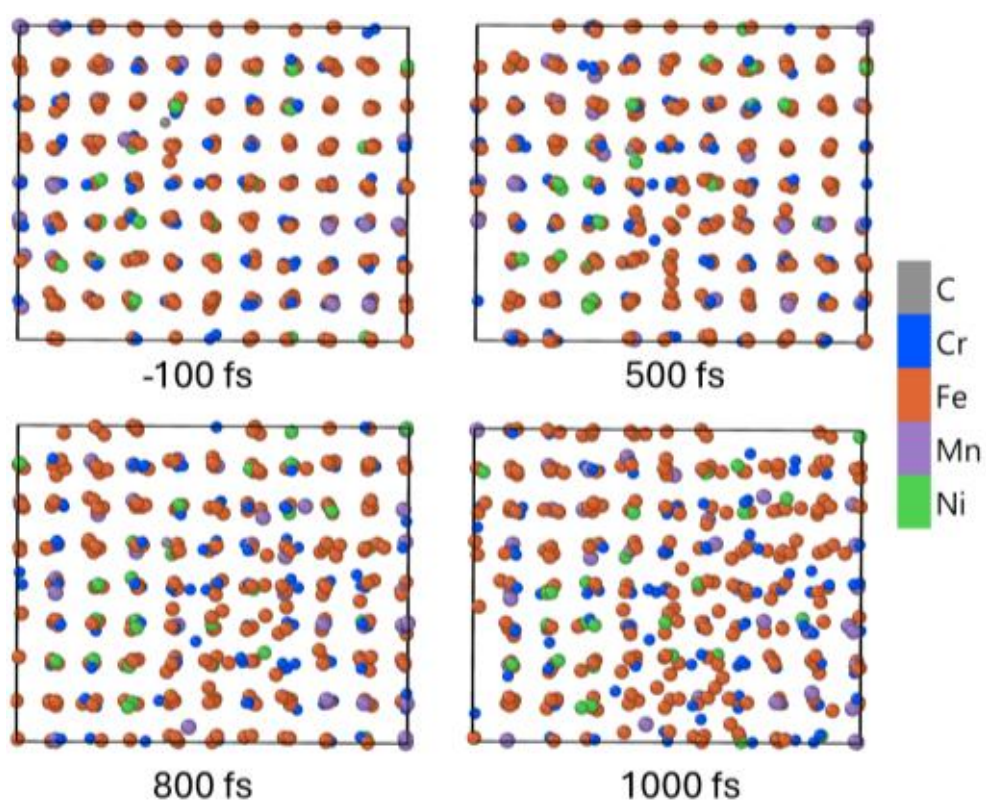

*Figure 9. Atomic snapshots of Born-Oppenheimer simulation (no electron-phonon coupling) of the supercell of stainless steel (fcc) after irradiation with 2 eV/atom absorbed dose, 10 fs FWHM, 30 eV photon energy, modeled with XTANT-3.*

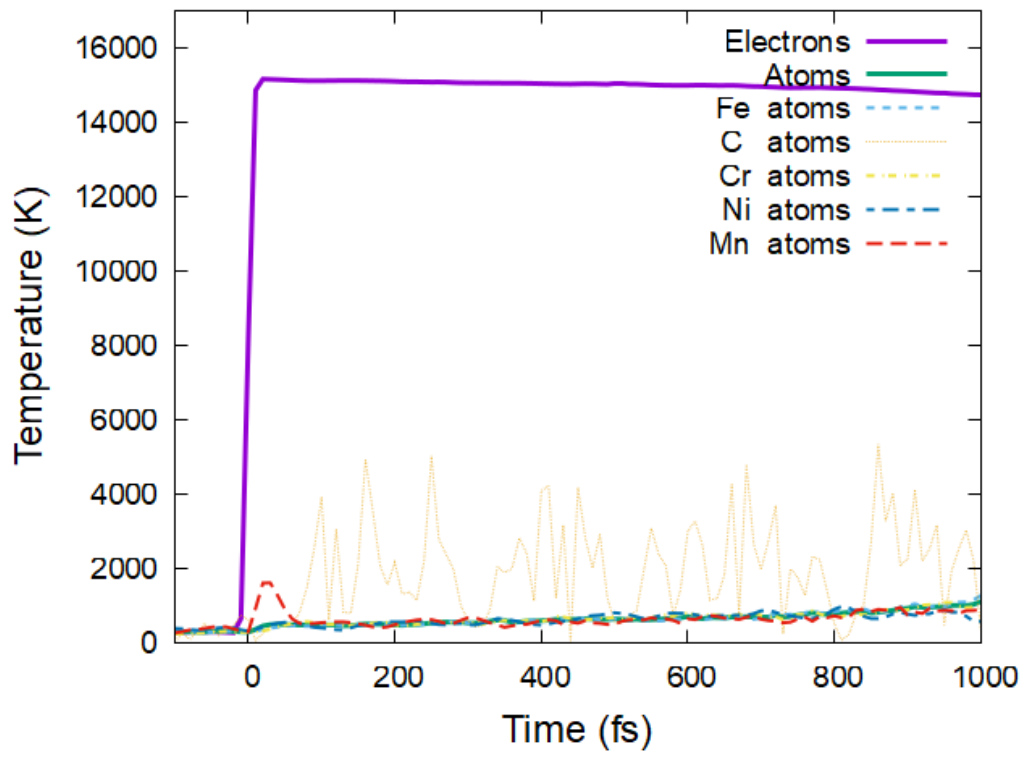

*Figure 10. Atomic and electronic temperatures in Born-Oppenheimer simulation (no electron-phonon coupling) of stainless steel (fcc) irradiated with 2 eV/atom absorbed dose, 10 fs FWHM, 30 eV photon energy, modeled with XTANT-3. Atomic temperatures of various species are also shown.*

The damage threshold doses can be converted into the incoming fluence thresholds using the photoabsorption length (see Appendix), assuming linear photon absorption, normal incidence of the laser pulse, and no energy transport or emission from the surface[15]. The thermal and nonthermal damage thresholds as a function of the photon energy are shown in Figure 11. Jumps in the curves correspond to particular shells of the elements, increasing the photoabsorption and thus reducing the damage threshold fluence at the corresponding photon energies [15]. These results may guide future experiments and help to estimate the material resilience under photon irradiation.

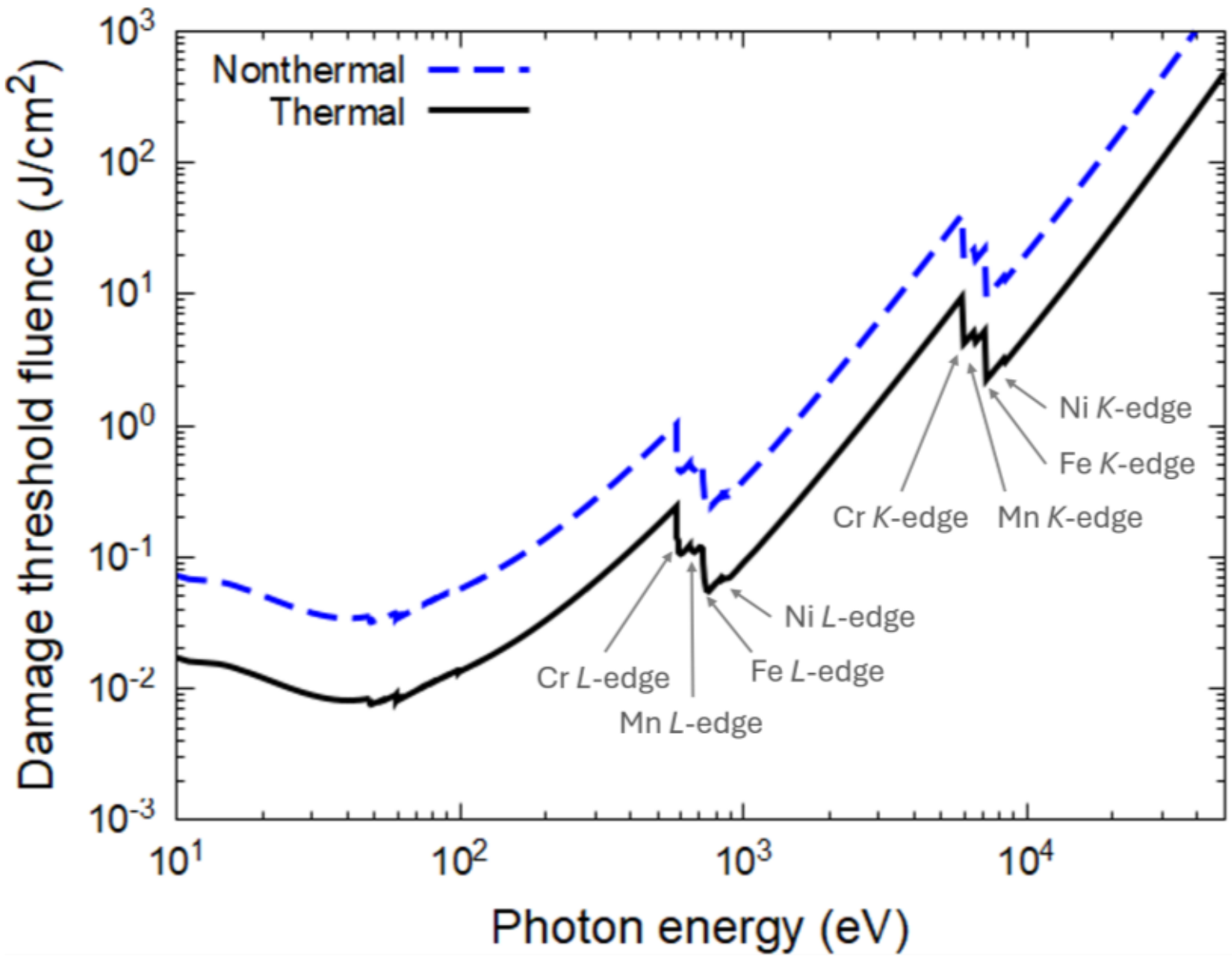

*Figure 11. Thermal and nonthermal melting threshold fluences in stainless steel estimated from the damage doses predicted with XTANT-3.*

# IV. Conclusions

Austenitic stainless steel ($Fe_{0.5875}Cr_{0.25}Mn_{0.09}Ni_{0.07}C_{0.0025}$) is simulated with the help of XTANT-3 hybrid code. It enabled us to evaluate the electronic heat capacity and the electronic heat conductivity up to the electronic temperatures of ~50,000 K. The electron-phonon coupling parameter was calculated using the nonperturbative dynamical coupling approach for the electronic temperature up to ~25,000 K. The damage threshold is estimated to be ~0.45 eV/atom, inducing thermal melting at the picosecond timescale.

It is also revealed that stainless steel has a nonthermal melting mechanism, which leads to atomic disorder solely *via* the electronic excitation modifying the interatomic potential energy surface (without requiring the atomic heating). Atoms, experiencing new forces in the electronically excited state, disorder at sub-picosecond timescale for deposited doses above ~1.9 eV/atom in a Born-Oppenheimer simulation (excluding the electron-phonon coupling). Atoms accelerate nonthermally during the induced phase transition. The manganese subsystem is especially susceptible to nonthermal acceleration due to electronic excitation.

The results indicate that a reliable simulation of stainless steel response to irradiation should account for the nonthermal effects, induced by modification of the interatomic potential in an electronically excited state. It is especially important at high doses, where the contribution of the nonthermal effects increases.

# V. Conflicts of interest

There are no conflicts to declare.

# VI. Data and code availability

The code XTANT-3 used to model X-ray irradiation is available from [30]. The tables with the calculated electronic heat capacity, conductivity, and electron-phonon coupling parameter are available from [59].

# VII. Acknowledgments

Computational resources were provided by the e-INFRA CZ project (ID:90254), supported by the Ministry of Education, Youth and Sports of the Czech Republic. The author thanks the financial support from the Czech Ministry of Education, Youth, and Sports (grant nr. LM2023068).

# VIII. Appendix

The photon attenuation length in the stainless steel, constructed from the atomic photoabsorption cross-sections [31] according to the stoichiometry of the material, is shown in Figure 12 (left panel). Note that this method does not account for collective effects in the valence/conduction band of the metal and thus cannot be extended to lower photon energies.

The total and partial electronic inelastic mean free paths in stainless steel, calculated with the complex-dielectric function formalism using the single pole approximation[32], are shown in Figure 12 (right panel). The formalism includes collective electron scattering (on plasmons) and thus describes the valence/conduction-band scattering.

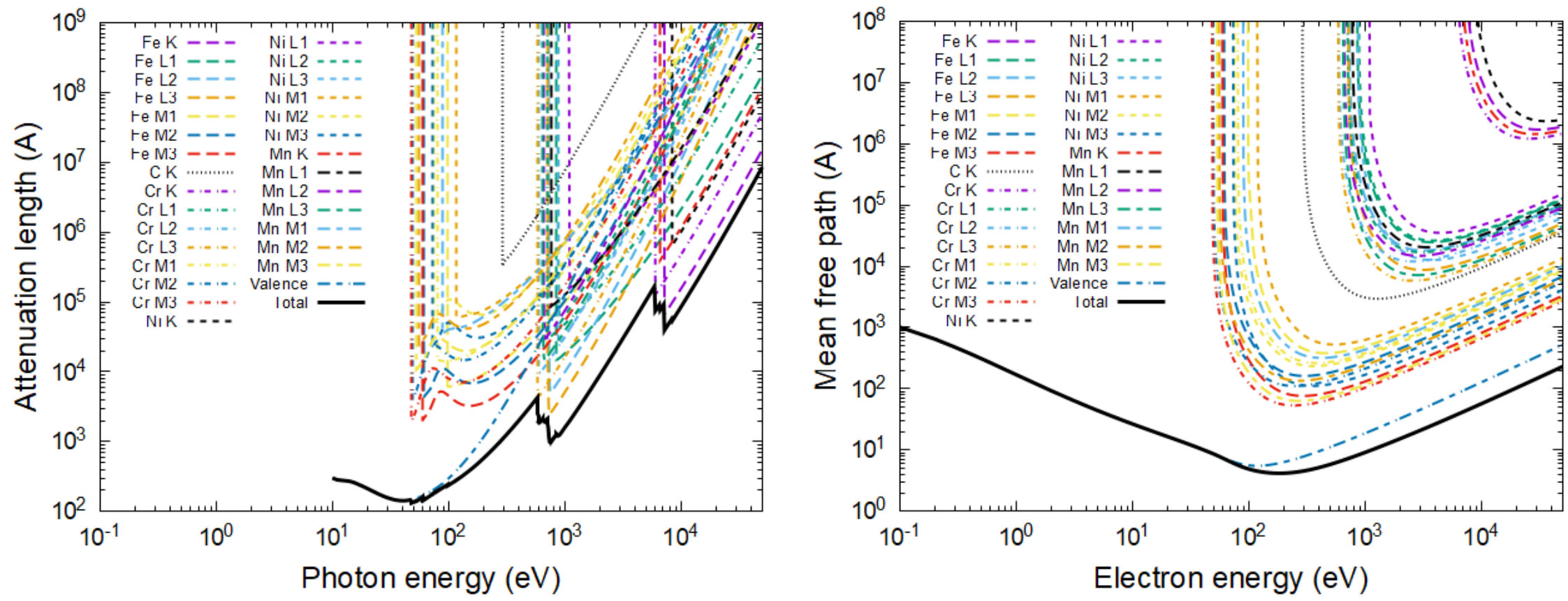

*Figure 12. Photon attenuation length (left panel) and electronic inelastic mean free path (right panel) in stainless steel. Total values are shown in black solid lines. Partial ones, for each atomic shell of each element, are dashed and colored.*